\documentstyle[aps,prl,epsf,twocolumn,epsfig,floats]{revtex}

\begin{document}
\draft


\wideabs{

\title{Anomalous electronic structure and pseudogap effects in Nd$_{1.85}$Ce$_{0.15}$CuO$_4$ }
\author{N.P. Armitage, D.H. Lu, C. Kim, A. Damascelli,
K.M. Shen, F. Ronning, D.L. Feng, P. Bogdanov,  and Z.-X. Shen}
\address{Dept. of Physics, Applied Physics and Stanford Synchrotron
        Radiation Laboratory, Stanford University, Stanford, CA 94305}
\author{Y. Onose, Y. Taguchi, and Y. Tokura}
\address{Department of Applied Physics, The
 University of Tokyo, Tokyo 113-8656, Japan}
\author{P.K. Mang, N. Kaneko, and M. Greven}
\address{Dept. of Applied Physics and Stanford Synchrotron
        Radiation Laboratory, Stanford University, Stanford, CA 94305}

\date{December 21, 2000}
\maketitle
\begin{abstract}

We report a high-resolution angle-resolved photoemission
spectroscopic (ARPES) study of the electron-doped ($n$-type)
cuprate superconductor Nd$_{1.85}$Ce$_{0.15}$CuO$_4$.  We observe
regions along the Fermi surface where the near-$E_F$ intensity is
suppressed and the spectral features are broad in a manner
reminiscent of the high-energy ``pseudogap'' in the underdoped
$p$-type (hole doped) cuprates.  However, instead of occurring
near the ($\pi, 0$) region, as in the $p$-type materials, this
pseudogap falls near the intersection of the underlying Fermi
surface with the antiferromagnetic Brillouin zone boundary.

\end{abstract}
} 

The electron-doped cuprate superconductors provide a unique
opportunity to study the physics of doped Mott insulators and
high-T$_c$ superconductors.  In addition to possessing interesting
physics in their own right, the $n$-type materials, with their
different normal state properties and phase diagram, offer an
alternative venue to test various theories of high-T$_c$
superconductivity.

Nd$_{2-x}$Ce$_{x}$CuO$_{4\pm \delta}$ is a member of the small family of 
cuprate superconductors that can be doped with electrons\cite{Tokura}.  
Only an approximate symmetry in the phase diagram exists about the zero 
doping line between $p$- and $n$-type, as the antiferromagnetic phase is 
much more robust in the electron-doped material and persists to much higher 
doping levels.  Superconductivity occurs in a doping range that is almost 
five times narrower.  In addition, these two ground states occur in much 
closer proximity to each other.  Experiments show other contrasting 
behavior between $n$-type superconductors and their $p$-type counterparts, 
for instance a T$^2$ dependence of the in-plane resistivity\cite{Gollnik}, 
the lack of incommensurate neutron scattering peaks \cite{Thurston} and a 
temperature induced sign reversal of the Hall coefficient\cite{Wang}.  
Although they appear to share a superconducting $d$-wave pairing symmetry 
\cite{Tsuei,NCCOGap}, the $2\Delta_{sc}/k_{B} T_{c}$ ratio of the 
n-type materials is much smaller than that of its optimally doped $p$-type 
counterparts \cite{Huang}.

Here we report a detailed angle resolved photoemission
spectroscopic (ARPES) study of the electronic structure of this
electron-doped cuprate superconductor.  Significant differences
are found between it and the $p$-type materials that shed light on
a number of important topics in the high-T$_c$ superconductors.
We find that the Luttinger volumed Fermi surface is truncated into
several pieces with a high-energy ``pseudogap''-like suppression
that forms, not at the maximum of the $d$-wave functional form as
in the underdoped p-type materials, but near the intersections of
the Fermi surface with the antiferromagnetic Brillouin zone (AFBZ)
boundary.

Single crystals of Nd$_{1.85}$Ce$_{0.15}$CuO$_4$ (NCCO) were grown
by the traveling solvent floating zone method in 4 ATM of O$_2$.
Details of this growth can be found elsewhere\cite{Onose}. The
resulting crystals show an onset of superconductivity at 25K and a
superconducting volume (Meissner shielding) of almost 100$\%$ at
20K.  A second batch of crystals, grown at Stanford University
under similar conditions, showed an onset at 24K with similarly
narrow transition widths. The photoemission data obtained between
the two batches are identical.

As it is imperative to determine which characteristics of the data are 
intrinsic and which are possible artifacts due to, for instance, the matrix 
elements for photoexcitation, data were collected on two different 
photoemission systems.  On beamline 5-4 at the Stanford Synchrotron 
Radiation Lab (SSRL), data were taken at 16.5 eV photon energy ($\hbar 
\omega$) with an incident angle of approximately 45 degrees and the 
in-plane polarization along the Cu-O bonds (Energy resolution $\Delta E$ 
$\simeq$10 meV and angular resolution $\Delta \theta$ $\simeq 
0.5^{\circ}$).  At beamline 10.0.1 at the Advanced Light Source (ALS) an 
incident energy of 55eV was used in a glancing incidence geometry with 
in-plane polarization at 45 degrees to the Cu-O bonds ($\Delta E$ $\simeq$ 
20meV, $\Delta \theta$ $\simeq 0.25^{\circ}$).  All displayed spectra were 
taken at low temperature ($10-20K$) in the superconducting state; this does 
not change the conclusions we make regarding the normal state electronic 
structure as the changes in spectra with the onset of superconductivity are 
very subtle in NCCO\cite{NCCOGap}.  The chamber pressure was lower than 4 x 
10$^{-11}$ torr.  Cleaving the samples at low temperature {\it in situ} 
results in shiny flat surfaces which low-energy electron diffraction (LEED) 
[Fig.  2(c)] reveal clean and well ordered with a symmetry commensurate 
with the bulk.  No signs of surface aging were seen for the duration of the 
experiment($\sim24$ hours).

In Fig.'s 1(a) - (c) we display energy distribution curves (EDCs)
from high symmetry lines in the Brillouin zone (BZ) taken on the
SSRL apparatus.  In Fig.  1(a), one sees a dispersion along the
$\Gamma$ to ($\pi, \pi$) direction which is ubiquitous among the
cuprates.  A broad feature disperses quickly towards the Fermi
energy ($E_F$), sharpens to a sharp peak at $\vec{k_F}$ ($0.46\pi,
0.46\pi$), and then disappears.  Along the $\Gamma$ to ($\pi, 0$)
direction, the $n$-type spectra are quite different from their
$p$-type counterparts.  While the low-energy feature in the
optimally doped $p$-type compounds disperses quite close to $E_F$
and forms a flat band region with a correspondingly high density
of states close to $E_F$, in NCCO this flat band region is at
about 300 meV higher energy\cite{King}.  Along ($\pi, 0$) to
($\pi, \pi$) [Fig.  1(c)] the peak disperses to $E_F$ and gives a
small peak at $\vec{k_F}$.  The spectra also possess a large
background contribution that has a weak maximum at 300 meV, as
seen in the 100$\%$ curves in Figs. 1(a) and 1(c).

In addition to the spectra from the high-symmetry directions, more
comprehensive intensity maps were obtained over a large region in
momentum space at both photon energies.  In Fig.  2(a) and 2(b) we
have plotted the integrated spectral weight of the EDCs ($\hbar
\omega$ = 16.5 and 55eV respectively) from within a 30 meV window
about $E_F$ as a function of $\vec{k}$.  This gives a measure of
the regions in momentum space that dominate the low-energy
properties of the material i.e.  the Fermi surface. The expected
large Fermi surface [as depicted in Fig. 2(d)]has a volume greater
than 1/2 (measured filling level $\delta =1.12 \pm
0.05$) and has a shape that is consistent with the LDA band
calculations\cite{Massida}. Interestingly however, it shows two
regions of suppressed spectral weight near ($0.65\pi, 0.3\pi$) and
($0.3\pi, 0.65\pi$).  Although there is obviously some modulation
of the intensity due to matrix elements, as some of the details of
the intensity pattern are different between the two excitation
energies, the gross features are the same.  In addition, data from
SSRL at 16.5eV with the sample turned so that the polarization is
at $45^{\circ}$ to the Cu-O bond reveals a similar pattern (not
shown).  The fact that the same systematics are seen in different
configurations gives us confidence that we are measuring intrinsic
properties, a fact which will be substantiated by lineshape
analysis below.

A detailed look at $\hbar \omega$ = 16.5eV EDCs through the suppressed 
region of the Fermi surface [Fig.  1(d)] reveals that the peak initially 
approaches $E_F$ and then monotonically loses weight despite the fact that 
its maximum never comes closer than $\sim$100 meV to $E_F$.  This results 
in features which are quite broad, even at $\vec{k_F}$.  Such behavior is 
similar to the high-energy pseudogap seen in the extreme underdoped 
$p$-type materials although in the present case it is observed near 
($0.65\pi, 0.3\pi$) and not at ($\pi, 0$) the maximum of the $d$-wave 
functional form.  An additional important difference is that a true clean 
gap at the very lowest energies cannot be defined, because the continually 
decreasing slope of the EDCs at $\vec{k_F}$ gives a leading edge midpoint 
that is within a few meVs of $E_F$.  We need to make the distinction 
between high-energy pseudogap behavior [the conversion of an expected sharp 
peak at $E_F$ to a broad feature over a large energy range ($\sim$100 meV) 
and the suppression of low-energy spectral weight] and low-energy pseudogap 
behavior which is the clean leading edge gap seen in the normal state of 
underdoped and optimally doped $p$-types near ($\pi, 0$) that has been 
posited to be reflective of pairing fluctuations in the normal state.  The 
low-energy pseudogap in the p-type materials appears to evolve somewhat 
continuously with underdoping into the high-energy pseudogap of the 
underdoped superconductor and AF insulator.  Both gaps can be distinguished 
in moderately doped samples.  The fact that they share the same $d$-wave 
functional form has lead to proposals that the pairing in the 
superconductor is an inherited property of the insulating state\cite{RVB}.  
These distinctions are important ones that we will return to later.

In Fig.  3a and 3b we plot $\hbar \omega$=55 and 16.5eV EDCs from
around $\vec{k_F}$ from ($\pi/2, \pi/2$) to ($\pi, 0.3\pi$) as
shown in the inset. Consistent with the intensity maps in Fig 2(a)
and 2(b) the data show a well defined and intense peak at $E_F$ in
momentum space regions close to the $\Gamma$ to ($\pi, \pi$) Fermi
surface crossing and the ($\pi, 0$) to ($\pi, \pi$) Fermi surface
crossing.  In between these regions, corresponding to where the
maps in Figs.  2(a) and 2(b) show suppressed intensity, the
spectra are rather different [see curves 4 of Fig.  3(a) and 7-9
of Fig.  3(c)] with no well defined peaks.  These qualitatively
different spectral lineshapes prove that the major effects seen in
the intensity plots are not due to matrix elements which will not
change the shape of the spectra drastically.  These assertions receive further
support from the measurement of EDC widths in Fig.  3(b).  Here we
plot the width in energy of the EDCs (defined as the FWHM/2 of the
actual features and not the output of a fitting parameter) in Fig.
3(a).  This quantity, which can be viewed as an average over the
low energy scattering rate (Im$\Sigma$) is largest in the momentum
region of interest along $k_F$ where the near-$E_F$ spectral
weight is suppressed maximally as shown in Fig.'s 2(a) and 2(b).
This analysis reveals those regions along the FS that show the
anomalous low-energy spectral weight suppression are subject to
the strongest scattering effects.

We take the view that this anomalous electronic structure in the
suppressed region is reflective of the transfer of spectral weight
from a near-$E_F$ peak to a large broader incoherent maximum at
higher energy.  As one moves along the Fermi surface from ($\pi/2,
\pi/2$) to ($\pi, 0.3\pi$), this high-energy part at first gains
intensity and shifts the global maxima to higher energy.  In the
intermediate region along the $\vec{k_F}$ contour, the near $E_F$
weight is suppressed maximally with a plurality of spectral weight
at higher energy.  Moving towards ($\pi, 0.3\pi$) the high-energy
part still forms a maximum at $\sim$100 meV but the low-energy
peak begins to recover at $E_F$ and the two features can be seen
simultaneously in a single spectrum.  The low-energy peak
continues to gain weight until the zone edge at $\sim$($\pi,
0.3\pi$), while the high-energy part loses intensity and
disappears.  The point of view that the two features are part of a
single spectral function is supported by the fact that we observe
only a single Fermi surface with the expected Luttinger volume.

We note here that the regions of momentum space with the unusual
low-energy behavior fall intriguingly close to the intersection of
the underlying FS with the AFBZ boundary as shown by the dashed
lines in Fig.  2(a) and 2(b). This suppression of low-energy
spectral weight and large scattering rate in certain regions on
the FS bears resemblance to various theories that emphasize a
coupling of charge carriers to a low-energy collective mode.
Central to these schemes is the supposition that there exists
anomalous low energy scattering channels with typically ($\pi,
\pi$) momentum transfer. A simple phase space argument such as
that shown in Fig.  2(d) shows that those charge carriers which
lie at the intersection of the FS with the AFBZ boundary will
suffer the largest effect of anomalous ($\pi, \pi$) scatterings as
these are the only FS locations that can have low-energy coupling
with Q = ($\pi, \pi$).  These heavily scattered regions of the FS
have been referred to in the literature as ``hot spots''.  It has
been suggested that the large back-scattering felt by charge
carriers in the hot spots is the origin of the pseudogap in the
underdoped hole-type materials\cite{Pines}.

A few candidates for such a scattering channel have been proposed.  One 
realization would be a strong coupling of the charge carriers to ($\pi, 
\pi$) magnetic fluctuations.  Indeed, strong AF fluctuations have been 
found at ($\pi, \pi$), as may have been expected in a material that is as 
close to the antiferromagnetic phase as this one\cite{Thurston}.  In this 
context the pseudogap is seen as a precursor to the formation of the 
antiferromagnetic phase.  Other possibilities include coupling to short 
range fluctuations of the CDW, DDW, or phononic type.  A great deal of 
theoretical effort has been devoted to schemes based on the above general 
considerations.\cite{Pines,Rice1,DDW}.  Of 
course we cannot rule out a combination of effects.  Alternatively, large 
effective ($\pi, \pi$) scatterings can be caused by umklapp (U) processes 
which are only allowed for charge carriers on the AFBZ boundary.  It has 
been suggested that these U processes create anomalously large effective 
backward scattering that leads to hot-spot phenomena and the creation of a 
non-Fermi liquid without an additional broken symmetry.\cite{Rice2}

In the underdoped $p$-type materials, due to band filling considerations, 
the underlying Fermi surface's area is smaller and hence its intersection 
with the AFBZ boundary is closer to ($\pi,0$).  This, along with the 
existence of the near $E_F$ ($\pi,0$) extended saddle points, would cause 
these regions of the Brillouin zone to feel the largest effect of ($\pi, 
\pi$) couplings if such processes exist.  As alluded to earlier, this may 
lead to the formation of a pseudogap whose momentum dependence closely 
mimics and can be confused with the functional form of the $d$-wave 
superconducting gap.  In the present case of NCCO, the underlying Fermi 
surface is more distant from the ($\pi,0$) regions and the band there is 
flat 300 meV below $E_F$.  In this scenario, the pseudogap regions would 
move away from ($\pi,0$) on the FS and along $\vec{k_F}$ towards 
($\pi/2,\pi/2$), as observed.

The class of theories that predict pseudogap formation at the
intersection of the AFBZ boundary with the FS contrast with those
that explain high-energy pseudogap and hot spot effects to result
from fluctuations of a very large energy superconducting pairing
scale that has been inherited from the AF insulator.  The latter
class of theories predict that the momentum dependence of the
pseudogap will follow the general trend of the $d$-wave
superconducting gap i.e.  attaining a maximum value at ($\pi,0$).
The advantage of the present experiments on the electron-doped
material is that the two different origins for pseudogaps give
different functional forms, whereas they will give similar ones in
the $p$-type materials as detailed above.  In the very underdoped
regime in the $p$-type materials, ARPES measures a high-energy
$d$-wave like pseudogap on the order of 100 meV near the ($\pi,0$)
region\cite{Marshall}.  As we observe a high energy pseudogap in
NCCO of a similar energy scale, but with a very different momentum
dependence, our results demonstrate the possibility that this very
large $d$-wave-like pseudogap in the extreme underdoped $p$-type
materials may not be related to preformed pairs, but instead may
be due to pre-emergent magnetic order or other similar phenomena.
Our observation still allows for the existence of a normal state
low-energy pseudogap (defined by the leading edge) that may be
related to the existence of pairing fluctuations above T$_c$.  In
this regard it appears that one must carefully discriminate
between high-energy pseudogap behavior, which may be caused by the
effects discussed above and is tied to the intersection of the FS
with the AFBZ boundary, and pairing fluctuations that may cause
pseudogap behavior on a lower energy scale and should follow the
$d$-wave functional form.  If one wishes to interpret the physics
of the $p$- and $n$-type cuprates in a comprehensive fashion then
our results suggest that some of the confusing pseudogap
phenomenology regarding energy scales and doping dependencies may
be reconciled by distinguishing effects of these kinds.

We thank X.J. Zhou and Z. Hussain for beamline support.  The Stanford 
Synchrotron Radiation Laboratory and the Advanced Light Source are both 
operated by the DOE Office of Basic Energy Science, Division of Chemical 
Sciences and Material Sciences.  Additional support comes from the Office 
of Naval Research: ONR Grants N00014-95-1-0760/N00014-98-1-0195.  The Tokyo 
crystal growth work was supported in part by Grant-in-Aids for Scientific 
Research from the Ministry of Education, Science, Sports, and Culture, 
Japan, and the New Energy and Industrial Technology Development 
Organization of Japan (NEDO).  The crystal growth at Stanford was supported 
by the U.S.  Department of Energy under Contract Nos.  
DE-FG03-99ER45773-A001 and DE-AC03-76SF00515 and by a NSF CAREER Award.  
M.G. is also supported by the A.P. Sloan Foundation.

\begin{figure}[htb]
\centerline{\epsfig{figure=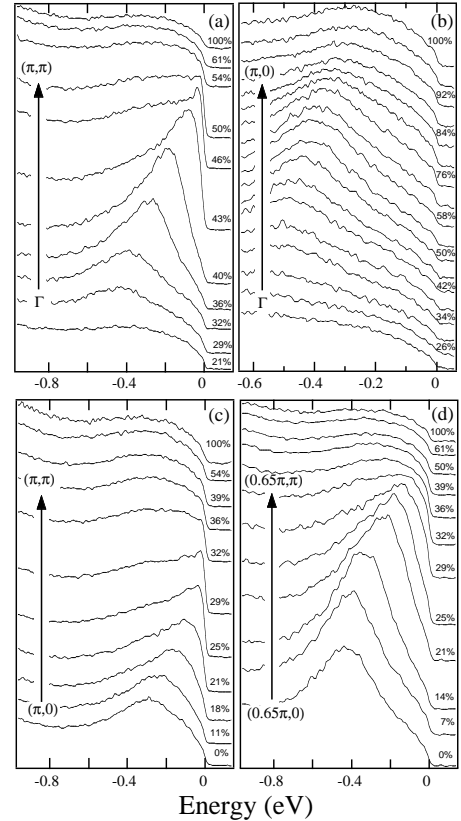,width=6cm}}
\vspace{.2cm} \caption{ Energy distribution curves from various
directions in the Brillouin zone taken with $\hbar \omega$ =
16.5eV at SSRL.  (a) $\Gamma$ - ($\pi, \pi$), (b) $\Gamma$ -
($\pi, 0$), (c) ($\pi, 0$)-($\pi, \pi$), (d) ($0.65\pi,
0$)-($0.65\pi, \pi$).  For a schematic refer to the black arrows
in Fig.  2(a).}
\end{figure}

\begin{figure}[htb]
\centerline{\epsfig{figure=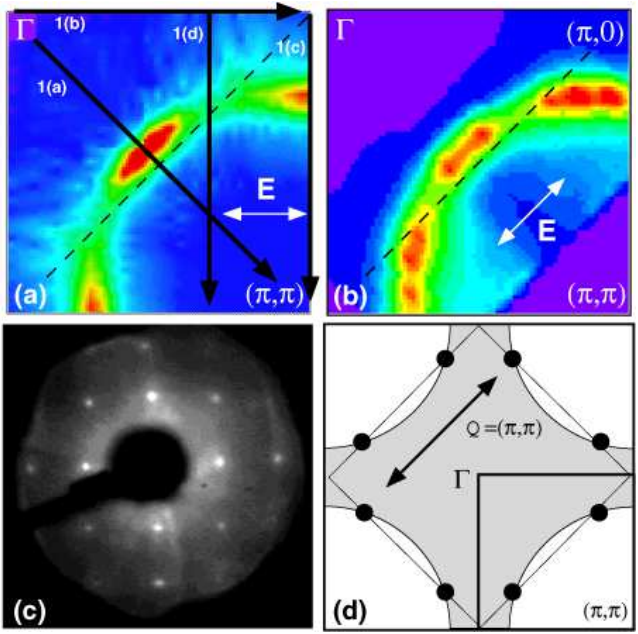,width=8.5cm}}
\vspace{.2cm}
 \caption{(color).  (a) and (b) Fermi surface of the partial Brillouin zone
 of NCCO taken with $\hbar \omega$ = 16.5 and 55eV respectively.  The plotted
 quantity is a 30 meV integration about $E_F$ of each EDC plotted as a
 function of $\vec{k}$.  16.5eV data were taken over a Brillouin zone octant
 and symmetrized across the $\Gamma$ to ($\pi, \pi$) line, while the 55eV
 data was taken over a full quadrant [11].  The polarization direction is
 denoted by the double ended arrow.  The dotted line is the
 antiferromagnetic Brillouin zone boundary. (c) LEED
 spectra of NCCO cleaved {\it in situ} at 10K. (d) Schematic showing only
 those regions of FS near the black circles can be coupled with a ($\pi,
 \pi$) scattering.}
\end{figure}

\begin{figure}[htb]
\centerline{\epsfig{figure=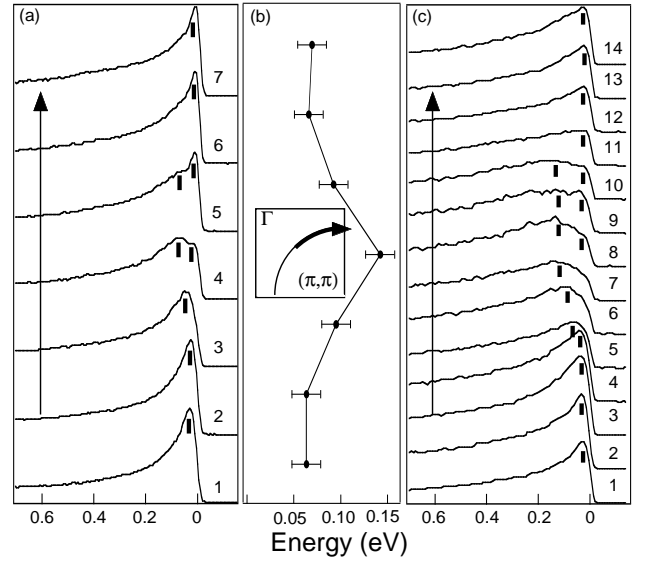,width=9cm}}
\vspace{.2cm} \caption{(a) EDCs from along the $\vec{k_F}$ contour for 
$\hbar \omega$=55 eV.  The graph plots ($\pi/2, \pi/2$) on the bottom and 
goes to ($\pi, 0.3\pi$) at the top along the $\vec{k_F}$ contour given in 
the inset.  (b) $\Delta E$ (defined as FWHM/2) for EDCs in Fig.  3(a).  (c) 
$\hbar \omega$=16.5 EDCs along the same $\vec{k_F}$ contour.  The large 
momentum independent background [defined as the signal at ($\pi, \pi$)] has 
been subtracted out.}

\end{figure}


\begin{references}
\bibitem{Tokura} Y. Tokura, H. Takagi, and S. Uchida,  Nature  {\bf 337}, 345-347 (1989)
\bibitem{Gollnik} F. Gollnik  and M. Naito, Phys. Rev. B {\bf 58}, 11734-11752 (1998)
\bibitem{Thurston} T.R. Thurston { \it et. al.}, Phys. Rev. Lett. {\bf 65}, 263-266 (1990);
K. Yamada { \it et. al.}, Journal of Physics and Chemistry of
Solids {\bf 60}, 1025-1029 (1999)
\bibitem{Wang} Z. Wang { \it et. al.} Phys. Rev. B, {\bf 43} 3020-3025 (1991);
S. Kubo, and M. Suzuki, Physica C, {\bf 185-189} 1251-1252 (1991);
W. Jiang { \it et. al.}, Phys. Rev. Lett. {\bf 73}, 1291-1293
(1994)
\bibitem{Tsuei} C.C.  Tsuei and J.R.  Kirtley, Phys.  Rev.  Lett.{\bf 85}, 
182-185 (2000); R.  Prozorov { \it et.  al.}, Phys.  Rev.  Lett.{\bf 85}, 
3700 (2000); J.  D.  Kokales { \it et.  al.}, Phys.  Rev.  Lett.,{\bf 85}, 
3696 (2000)
\bibitem{NCCOGap} N.P. Armitage { \it et. al.},  Phys. Rev. Lett. {\bf 86}, 1126 (2001)
\bibitem{Huang} Q. Huang { \it et. al.},    Nature, {\bf 347} 369-372 (1990)
\bibitem{Onose} Y. Onose { \it et. al.},  Phys. Rev. Lett. {\bf 82} 5120 (1999)
\bibitem{King} D. King  { \it et. al.},   Phys. Rev. Lett. {\bf 70}, 3159-3162 (1993);
R.O. Anderson  { \it et. al.}  Phys. Rev. Lett., {\bf 70},
3163-3166 (1993)
\bibitem{Disclaimer} A line of symmetrized data points was used
immediately below and parallel to the zone diagonal as these
points were missing from our initial data set.
\bibitem{Massida} S. Massida, N. Hamada, J. Yu, and A. Freeman, Physica C  {\bf 157}, 571-574 (1989)
\bibitem{RVB} G.  Kotliar and J.  Liu, Phys.  Rev.  B {\bf 38}, 5142 
(1988); H.  Fukuyama, Prog.  Theor.  Phys.  Suppl.  {\bf 108}, 287 (1992); 
P.  A.  Lee and X.  G.  Wen, Phys.  Rev.  Lett.  {\bf 76}, 503 (1996)
\bibitem{Pines} D. Pines, Z. Phys. B {\bf 103}, 129 (1997); J. Schmalian,
D. Pines and B. Stojkovic, Phys. Rev. Lett. {\bf 80}, 3839 
(1998); A.P. Kampf and J.R. Schrieffer, Phys. Rev. B {\bf 42}, 7967 (1990)
\bibitem{Rice1} R. Hlubina and T.M. Rice, Phys. Rev. B {\bf 51}, 9253 (1995)
\bibitem{DDW} S. Chakravarty { \it et. al.}, Phys. Rev. B {\bf 63}, 094503 (2001)
\bibitem{Rice2} N. Furukawa, T.M. Rice, and M. Salmhofer, Phys. Rev. Lett.
{\bf 81}, 3195 (1998);  C. Honerkamp{ \it et. al.}, Phys. Rev. B
{\bf 63}, 035109 (2001)
\bibitem{Marshall} D.S. Marshall { \it et. al.}, Phys. Rev. Lett. {\bf 76}, 4841 (1996)

\end{references}
\end{document}